\documentclass [12pt,a4paper]{article}
\usepackage{times}

\DeclareFontFamily{OT1}{times}{}
\DeclareFontShape {OT1}{times}{m }{n }{ <-> ptmr }{}
\DeclareFontShape {OT1}{times}{bx}{n }{ <-> ptmb }{}
\DeclareFontShape {OT1}{times}{m }{it}{ <-> ptmri}{}
\DeclareFontShape {OT1}{times}{bx}{it}{ <-> ptmbi}{}
\usepackage{amsmath}
\usepackage{amsfonts}
\usepackage{amssymb}
\usepackage{latexsym}
\newcommand{\conj}{\overline}         % conjugate
\newcommand{\Scal}{\mathbb{S}}        % scalar part unary  S[AB] = s
\newcommand{\cl}{C \kern -0.1em \ell} % Clifford algebra
\begin{document}
\title{\vspace{-1cm} \bf Lanczos's equation to replace\\ Dirac's equation~?}

\author{Andre Gsponer and Jean-Pierre Hurni\\ ~\\ \emph{Independent Scientific Research Institute}\\ \emph{Box 30, CH-1211 Geneva-12, Switzerland.}}

\date{\normalsize Published {\it in} J.D. Brown \emph{et al.}, eds.,  Proceedings of the\\ Cornelius Lanczos International Centenary Conference\\ Raleigh, North Carolina, December 12--17, 1993\\ (ISBN 0-89871-339-0, SIAM, Philadelphia, 1994) 509--512.}
\maketitle
\begin{abstract}
     Lanczos's quaternionic interpretation of  Dirac's  equation provides  a  unified description for all elementary particles  of spin  0,  $\frac{1}{2}$,  1, and $\frac{3}{2}$.   The Lagrangian formulation given  by  Einstein  and  Mayer  in  1933  predicts  two  main  classes   of solutions. ~(1)  Point  like partons which come in two  families, quarks and leptons.  The correct fractional or integral  electric and baryonic charges,  and zero mass for the neutrino and the  u-quark,  are  set  by  eigenvalue  equations.   The  electro-weak interaction  of  the  partons is the same as  with  the  Standard model,  with the same two free parameters:  $e$ and $\sin^2 \theta$. There is no need  for a Higgs symmetry breaking mechanism. ~(2)  Extended hadrons for which there is no simple eigenvalue equation for  the mass.  The strong interaction is essentially non-local.  The pion mass  and pion-nucleon coupling constant determine to first order the nucleon  size, mass and anomalous magnetic moment.

\end{abstract}

     Since  1928,  Dirac's  relativistic wave-equation  has  been rewritten   in  various  forms   in  order  to   facilitate   its interpretation.   For example,  in 1930, Sauter \cite{1} and Proca \cite{2} rewrote it using Clifford numbers.  However, the most direct road was taken  by  Lanczos in 1929 \cite{3}.  He showed how to derive  Dirac's equation from the coupled biquaternion system:
\begin{equation}\label{E1}
                 \conj{\nabla} A = m B, ~~~~~~~ \nabla B = m A.       % (1)
\end{equation}
     As   Lanczos's dissertation  was  an  attempt  to  formulate classical  electrodynamics as a quaternion field theory  \cite{4},  he knew that Maxwell's equations could be written:
\begin{equation}\label{E2}
                 \conj{\nabla} A = m B, ~~~~~~~ \nabla B = 0.         % (2)
\end{equation}
Hence,  \eqref{E1}  can  be seen as Maxwell's equations with  feed-back.  This was a very important idea,  and precisely the one  that lead Proca  in 1936 to discover the correct equation for  the  massive spin one particle \cite{5}.   The concept was easily generalized,  and Kemmer  finally  wrote  down  the  wave-equations  of  the (pseu\-do-)scalar  and  (pseudo-)vector  particles in the  form  we still  use  today  \cite{6}.  Later,  G{\"u}rsey \cite{7} showed that  Proca's  and Kemmer's equations are just degenerated cases of Lanczos's.

     In fact,  Lanczos's equation also admits spin $\frac{3}{2}$  solutions. These  are apparently devoid of the problems which  plague  the usual  formulations of spin  $\frac{3}{2}$.  Thus,  \eqref{E1}  provides  a  unified description  of all elementary particles,  showing that in  their fundamental state their spin must be 0, $\frac{1}{2}$, 1, or $\frac{3}{2}$.

     In the standard spinor or two-component formalisms,  Dirac's equation can be written as follows:
\begin{equation}\label{E3}
                   \partial L = m R,  ~~~~~~~  \partial R = m L.        % (3)
\end{equation}
Here $L$ and $R$ and the left- and right-handed components of Dirac's four-component spinor.   The main difference between \eqref{E1} and  \eqref{E3} is  that  $A$ and $B$ are biquaternions with four complex  components while  $L$ and  $R$ have only two.   This is the  `doubling'  problem that puzzled Lanczos for many years.  It often reappears, e.g., in relativistic  Hamiltonian dynamics \cite{8} or general relativity \cite{9}, whenever Dirac's equation emerges in a form or another.

     To  get Dirac's spin $\frac{1}{2}$ field,  Lanczos made the  following superposition:
\begin{equation}\label{E4}
                       D =  A \sigma  +  B^* \conj{\sigma}.          % (4)
\end{equation}
 Here  $\sigma$ is an idempotent\footnote{We assume $\sigma=\frac{1}{2}(1+i\vec{e}_3)$ where $\vec{e}_3$ is the third quaternion unit.} which has the effect of  projecting out  half  of  $A$,  which  added to another half  of  the  complex conjugated  of $B$,  gives a Lorentz covariant  superposition  that obeys the following wave equation:
\begin{equation}\label{E5}
                  \conj{\nabla} D = m D^* i\vec{e}_3.            % (5)
\end{equation}
This  equation,  to  be  called the  \emph{Dirac-Lanczos  equation},  is equivalent to Dirac's equation.   It will be rediscovered by other people, in particular by G{\"u}rsey \cite{7} and Hestenes \cite{10}.

     An important observation is that one can go from \eqref{E1} to  \eqref{E3} by  simply requiring that $A$ and $B$ are singular quaternions  \cite{11}.  However,  Dirac's  system \eqref{E3} does not only involve half as many components  as  Lanczos's system \eqref{E1},  it also  incorporates  the ingredients that make fermions essentially different from bosons, a  feature  that is embodied in the Lanczos-Dirac  equation  \eqref{E5}.  In effect, studying the time-reversal transformation, one finds that $\mathcal{T}^2=-1$, which means that \eqref{E5} implies the Pauli exclusion principle and Fermi statistics \cite[p.\,41]{12}.

       The origin of the difference between fermions and bosons  is to  be  found in the superposition \eqref{E4} which involves  a  complex conjugation and singles out a unique, but arbitrary, direction in space, i.e., $\vec{e}_3$ when $\sigma=\frac{1}{2}(1+i\vec{e}_3 )$.  Then, when the operators 
\begin{equation}\label{E6}
                            \mathcal{P} = (~)^*_{\{ -\vec x\}},
                     ~~~~~~ \mathcal{T} = (~)^*_{\{ -t\}}i\vec{e}_2,
                     ~~~~~~ \mathcal{C} = (~)i\vec{e}_2,              % (6)
\end{equation}
are applied to plane waves such as $D$=$D_0\exp[\vec{e}_3(Et$-$\vec{p}$$\cdot$$\vec{x}\,)]$,  one  sees that Lanczos's theory  necessarily implies the Stueckelberg-Feynman interpretation of  antiparticles.  Indeed,  since  $\vec{e}_2$  anticommutes with $\vec{e}_3$,  the effect of $\mathcal{C}$ is to reverse the sign of the argument in the exponential.  Moreover, these three involutions satisfy the CPT theorem, i.e., $\mathcal{C} \mathcal{P} \mathcal{T} = \mathbf{1}(~)_{\{-t,-\vec x\}}$, and finally $\mathcal{T}^2=-1$ as required for spin~$\frac{1}{2}$.

     The  four  rest-frame  solutions,  spin  up-down,  particle-antiparticle  are:  $D_0  = 1,  \vec{e}_1,  i\vec{e}_3,$  and $i\vec{e}_2$.   Operating by $\mathcal{I}=(~)i\vec{e}_1$,  one obtains another quartet of solutions which  together with  the first one make the spin $\frac{3}{2}$ solutions. 

      However,  the superposition \eqref{E4} is not the only one leading to  a  spin $\frac{1}{2}$ field obeying equation \eqref{E5}.   As shown  by  G{\"u}rsey in 1957, if \eqref{E4} represents a proton, the neutron is then \cite{13}:
\begin{equation}\label{E7}
           N = ( A\conj{\sigma} - B^* \sigma ) i\vec{e}_1.              % (7)
\end{equation}
In  particular,  Nishijima's  operator  $\exp{[\sigma i\varphi(t,\vec{x}\,)]}$,  \cite{14}, which leaves  the  $N$  field invariant,  shows that  $D$ is  electrically charged while $N$ is neutral.  Hence,  Lanczos's doubling is nothing but \emph{isospin}.   G{\"u}rsey's articles \cite{13,15} had a tremendous  impact \cite{16} to \cite{22}.   They  demonstrated  that `internal' symmetries such  as isospin  are  explicit and trivial in Lanczos's formulation  \eqref{E1}, while only space-time symmetries are explicit in \eqref{E5}.

     When he wrote his 1929 papers on Dirac's  equation,  Lanczos was with Einstein in Berlin.  In 1933, Einstein and Mayer (using semi-vectors,  a  formalism allied to quaternions) derived a spin $\frac{1}{2}$  field  equation (in fact,  a generalized  form of Lanczos's equation)  predicting  that particles would come in doublets  of different  masses  \cite{23}.   The  idea was that  the  most  general Lagrangian for quaternionic fields,  to be called the \emph{Einstein-Mayer-Lanczos (EML) Lagrangian}, should have the form:
\begin{equation}\label{E8}
 L = \Scal \Bigl[ A^+\conj{\nabla}A + B^+{\nabla}B  - (A^+ BE^+ + B^+ AE ) + (...)^+  \Bigr].  % (8)
\end{equation}
The field equations are then
\begin{equation}\label{E9}
             \conj{\nabla} A = B E^+,  ~~~~~~~ \nabla B =  A E,       % (9)
\end{equation}
which reduces to \eqref{E1} when $E=m$.    In the general case, the second order  equations for $A$ or $B$ become eigenvalue equations for  the mass.  The  generalization \eqref{E9} is obtained by the substitution $Am \rightarrow AE$ in the Lagrangian leading to \eqref{E1}.  Therefore, mass-generation is linked to a maximally parity violating field.

     There  are  two basic conserved  currents,  the  probability current $C$, and the barycharge current $S$:
\begin{equation}\label{E10}
     C= AA^+   + \conj{BB^+}, ~~~~~~ S= AEA^+   + \conj{BEB^+}.    % (10)
\end{equation}
Keeping  $E$  constant,  $C$ is invariant in any non-Abelian  unitary gauge transformation, $SU(2) \otimes U(1)$, of $A$ or $B$.   On the other hand,  $S$  is only invariant for Abelian gauge transformations  which also  commute  with $E$:  this is the general Nishijima group  \cite{14} which contains the electric and baryonic gauge groups.

     Of  special  interest  are the cases in which $E$ is  also  a global gauge field. The first such gauge is when $E$ is idempotent. One  solution of \eqref{E9} is then massive and the other one  massless \cite{24}. The most general local gauge transformations compatible with \eqref{E8} are elements of the unitary Nishijima group  ${U_N}(1,\mathbb{C})$ combined with \emph{one} non-Abelian gauge  transformation  which  operates on $E$ and $A$,  or $E$ and $B$, exclusively.   This  leads  directly  to the  Standard  model  of electro-weak interactions.

     A remarkable property of the Nishijima group is that it has exactly two non-trivial decompositions in a product $U(1,\mathbb{Q}) \otimes U(1)$, such  that  the electric charge is  quantized.  The  doublet  has charges $(0,-1)$ in one,  and $(+\frac{2}{3},-\frac{1}{3})$ in the other, $U(1,\mathbb{Q})$ being the discrete ring $R_4$ or $R_6$, respectively.  Hence, electron charge quantization necessarily implies the existence of quark states of fractional  electric  charge,  with both the neutrino and the  u-quark massless.

     The other fundamental case is when $E$ is real: $E$ describes a nucleonic field, non-locally coupled to a pseudoscalar eta-pion field.  The chiral $SU(2) \otimes SU(2)$ symmetry of low-energy strong interaction is explicit because one can perform independent   isotopic   rotations   on  the $A$ and  $B$ fields independently.  From this stage,  by various transformations  and additions,  one can easily get the sigma-model \cite{19}, Heisenberg's \cite{21} or Nambu-Jona-Lasinio's \cite{22} non-linear theories, the Skyrme model \cite{25}, etc.

     Lanczos's  and  Einstein-Mayer's theories provide a  unifying framework  for  electro-weak and low-energy strong interactions.  The  gauge invariance  of  the {EML} Lagrangian leads to a  surprising  formal proximity  of  the W-bosons in the electro-weak sector  with  the pions in the strong interaction sector.  Of  course,  differences between electro-weak and strong interactions are  considerable. But, leaving the details aside, any comparison should  involve `$e$' on one side,  and the pseudoscalar pion-nucleon  coupling constant `$g$' on the other.  That this is the case is well known.   For example,  the masses of nucleons and electrons are such that:
\begin{equation}\label{E11}
                 \frac{g^2}{e^2}   \approx \frac{M}{m}.       % (11)
\end{equation}

     What  remains to be studied,  is whether all of this is just an approximation to some more fundamental theory,  or whether the quark model and QCD will come out as high energy approximations in a biquaternion field theory of elementary particles,  an  idea suggested  by Lanczos for electrons and protons in his PhD thesis of 1919.

~\\
\noindent{\bf \large Acknowledgments} 
% ----------------------------------

     We wish to express our gratitude to Professor J.A.\ Wheeler, Professor  W.R.\ Davis,  and  Professor G.\ Marx  for  their  kind encouragement.

~\\
\noindent{\bf \large References} 
% ------------------------------

\newpage

\noindent{\bf \large Errata} 
% --------------------------

This electronic version corrects a number of typographical errors that plagued the published paper version.

The restriction of the Nishijima group to $U(1,\mathbb{Q}) \otimes U(1)$ is not sufficient to set the correct fractional or integral electric and baryonic charges of the quarks and leptons. The statement in the abstract \emph{``The correct fractional or integral  electric and baryonic charges,  and zero mass for the neutrino and the  u-quark,  are  set  by  eigenvalue  equations''} is thus wrong.  

Moreover, the spin 3/2 interpretation of Lanczos's equation is not as simple as suggested in the text: see arXiv:math-ph/0210055.

\end{document}